\documentclass[10pt, twocolumn]{IEEEtran}
\usepackage{epsfig,latexsym}
\usepackage{float}
\usepackage{indentfirst}
\usepackage{amsmath}
\usepackage{bm}
\usepackage{amssymb}
\usepackage{times}
\usepackage{graphicx}
\usepackage[ruled,linesnumbered,boxed]{algorithm2e}
\usepackage[noend]{algpseudocode}
\usepackage{subfigure}
\usepackage{psfrag}
\usepackage{hyperref}
\usepackage{cite}
\usepackage{lastpage}
\usepackage{fancyhdr}
\usepackage{color}
 \usepackage{amsthm}
 
\usepackage{bigints}
\newcounter{problem}
\newcounter{save@equation}
\newcounter{save@problem}
\makeatletter

\usepackage{changes}

\usepackage{comment}
\usepackage[normalem]{ulem}

\DeclareRobustCommand{\Erase}{\bgroup\markoverwith{\textcolor{red}{\rule[.5ex]{2pt}{0.4pt}}}\ULon}

\begin{document}
\title{{\LARGE One-hot Coding-based URA with RFFI-Enabled Message Authentication}}

\author{Wenbo Fan, Zeping Sui, \IEEEmembership{Member, IEEE}, Yuhei Takahashi, Jun Cheng, \IEEEmembership{Member, IEEE}, \\Zilong Liu, \IEEEmembership{Senior Member, IEEE}, and Pingzhi Fan, \IEEEmembership{Life Fellow, IEEE} 
\vspace{-3em}
\thanks{W. Fan and P. Fan are with the Information Coding  \& Transmission Key Lab of Sichuan Province, CSNMT Int. Coop. Res. Centre (MoST), Southwest Jiaotong University, Chengdu, China. W. Fan is also with the School of Computer Science and Electronic Engineering, University of Essex, Colchester CO4 3SQ, U.K. (e-mails: \href{mailto:fwb@my.swjtu.edu.cn}{fwb@my.swjtu.edu.cn}; \href{mailto:pzfan@swjtu.edu.cn}{pzfan@swjtu.edu.cn}).}
\thanks{Z. Sui and Z. Liu are with the School of Computer Science and Electronic Engineering, University of Essex, Colchester CO4 3SQ, U.K. (e-mails: \href{mailto:zepingsui@outlook.com}{zepingsui@outlook.com}; \href{mailto:zilong.liu@essex.ac.uk}{zilong.liu@essex.ac.uk}).}
\thanks{Y. Takahashi and J. Cheng are with the Department of Intelligent Information Engineering and Science, Doshisha University, Kyoto 610-0321, Japan (e-mails: \href{mailto:cyjk1101@mail4.doshisha.ac.jp}{cyjk1101@mail4.doshisha.ac.jp}; \href{mailto:jcheng@ieee.org}{jcheng@ieee.org}).}}

 \maketitle

\vspace{0cm}

\begin{abstract}
Unsourced random access (URA) has emerged as a promising paradigm for enabling massive connectivity in Internet-of-Things (IoT) networks. However, since URA transmissions generally don't carry any device identifiers, the receiver may not associate decoded messages with their originating devices, introducing a security vulnerability: forged messages may be decoded as legitimate. To address this problem, we present a one-hot coding (OHC)-based URA framework that enables message authentication while preserving the unsourced transmission principle. Specifically, distinct messages are mapped onto orthogonal channel uses via an OHC-based common codebook and transmitted using on-off keying modulation. The resulting orthogonal channel structure enables radio-frequency fingerprint identification to authenticate received signals by exploiting device-specific hardware impairments. Analytical expressions for the per-user probability of error and the probability of successful spoofing are derived. Numerical results demonstrate that the proposed scheme enables secure URA transmission while maintaining reliable communication performance in ultra-short-payload IoT scenarios.

\end{abstract}

\vspace{-1.3em}
\section{Introduction}
Future wireless networks are expected to support efficient and secure information exchange among a massive number of machine-type devices \cite{10745245}, such as sensor data transmission in smart city or smart factory applications, where the payload is often extremely small (e.g., around 10 bits), enabling ultra-short-payload uplink transmission in massive-access settings with security requirements.

Unsourced random access (URA) was first proposed in \cite{8006984} to support massive connectivity, whereby all devices share a common codebook and transmit messages without identifiers, and the BS recovers messages from the received signal without associating them with their originating devices \cite{11269777}. 

The performance metric for URA is typically characterized by the minimum $E_b/N_0$ required to support $K_a$ active devices while meeting a target per-user probability of error (PUPE). An achievability bound on this metric, accounting for \textit{finite-blocklength} effects in a Gaussian channel, was derived in \cite{8006984}. Existing URA schemes attaining this bound fall into three categories: $T$-fold ALOHA, segmentation coding, and two-phase coding. In $T$-fold ALOHA \cite{8006985}, the receiver recovers up to $T$ messages per slot; combined with irregular repetition slotted ALOHA (IRSA) \cite{10266359}, the resulting $T$-fold IRSA scheme \cite{8830448} still exhibited a $5$--$10$ dB gap to the bound. In segmentation coding, each message is divided into sub-blocks recovered via compressed sensing (CS) and stitched together using tree-based coded compressed sensing (CCS) \cite{9153051} and dynamic compressed sensing (DCS) \cite{10535261}, the latter narrowing the gap to $1$--$3$ dB. \textcolor{blue}{In two-phase coding, each message is split into two parts: the first part is transmitted via a CS-based URA scheme\cite{9797778}, and the second part is transmitted via on-off division multiple access (ODMA) \cite{11006717,10415634,10105449,11677275,11657329}.}

Although these schemes have narrowed the gap to the achievable bound, their unsourced nature introduces security challenges: spoofing attacks become possible once the common codebook is compromised. A key-based authentication method was proposed in \cite{9564037} from a cryptographic perspective. However, its additional authentication payload and cryptographic processing increase both communication overhead and computational complexity. Since IoT devices are typically low-cost and powered by batteries or energy harvesting \cite{10980490}, lightweight authentication with low overhead is highly desirable. Radio-frequency fingerprint identification (RFFI) is a promising physical-layer authentication technique that exploits transmitter-specific hardware impairments without requiring an authentication payload \cite{10978824}. However, applying RFFI to URA is nontrivial, because reliable fingerprint extraction generally requires observing a signal from a single transmitter \cite{11369874}. We refer to this condition as the \emph{single-transmitter signal
observation (STSO)} requirement. In existing URA schemes \cite{8830448,9153051,10535261,9797778,11006717,10415634,10105449,11677275,11657329}, simultaneous transmissions over shared resources result in signal superposition, making direct application of RFFI difficult.

\textcolor{blue}{Against this background, we propose an authenticated URA framework based on one-hot coding (OHC), in which the $B$-bit message is mapped to a one-hot codeword of length $2^B$ and transmitted via on-off keying (OOK). This construction can be viewed as a special case of ODMA. Unlike existing ODMA-based URA schemes, the proposed scheme transmits the entire message without partitioning it into two segments and activates only one channel use per message.} Each collision-free active channel use then contains a signal from only one transmitter and satisfies the STSO requirement, thereby enabling RFFI-based authentication. Device identity is used internally at the receiver solely for physical-layer authentication and is not explicitly embedded in the transmitted URA payload.

\begin{figure}[t!]\centering \vspace{-0em}
	 {\epsfig{file=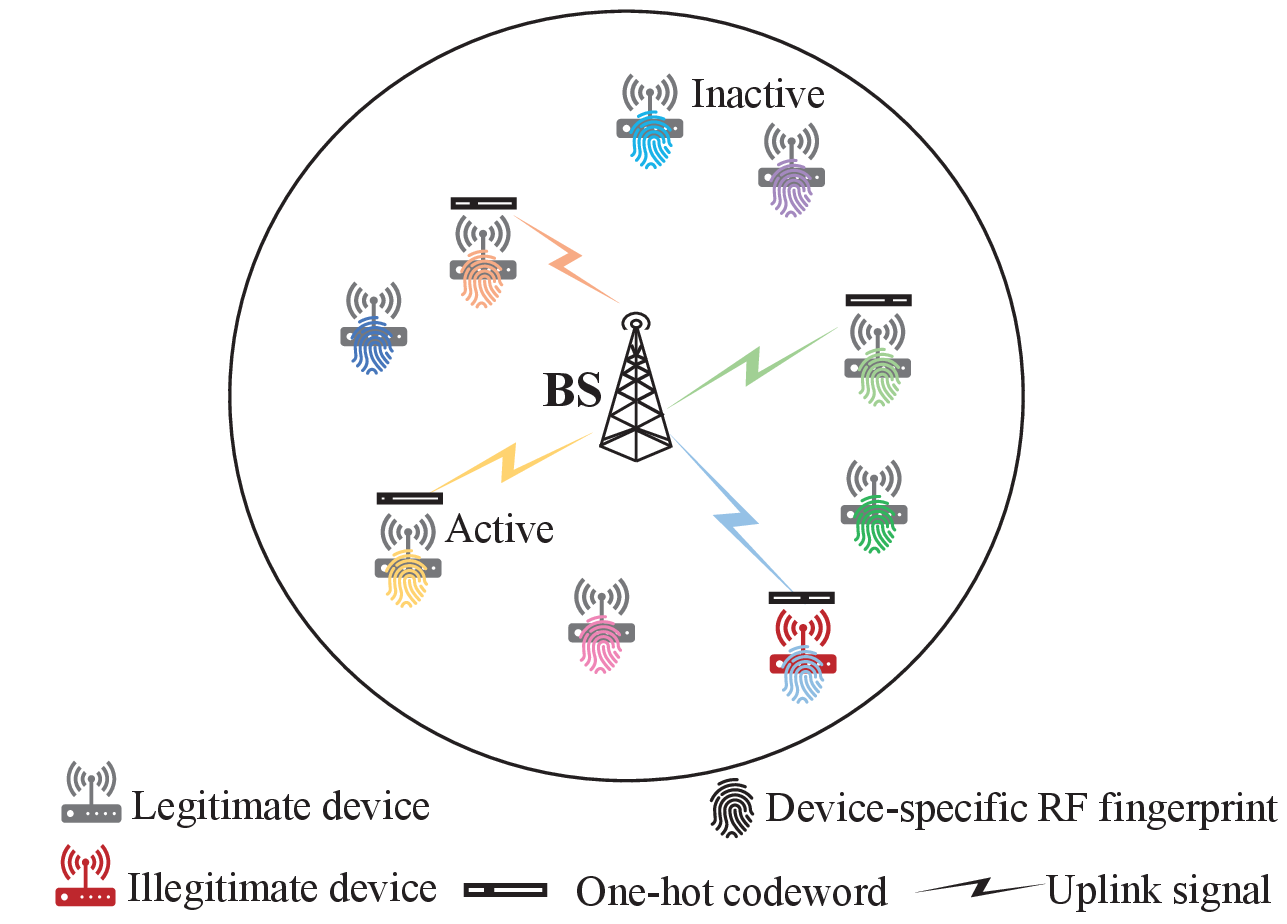, width=0.35\textwidth, clip=}}\vspace{-0.5em}
	\caption{ An illustration of the sparse activity of devices with device-specific RF fingerprints.
		\vspace{-1em}    }\label{scheme0}   \vspace{-1em} 
\end{figure}
\vspace{-1em} 

\section{System Model and Proposed Scheme} \label{section 2}

\subsection{Proposed Secure URA System}
We consider a wireless sensor network in which a large number of devices aim to transmit short messages to the BS under the URA protocol \cite{8006984}. As shown in Fig. \ref{scheme0}, at any given time, only a small subset of devices is active in each uplink transmission round. Since active devices transmit messages without identifying information, illegitimate devices may inject forged messages that the BS may decode as legitimate, thereby posing potential security threats. To address the above security issue, we distinguish between legitimate and illegitimate devices in the network. A device is considered legitimate if it is registered in the BS's access list; otherwise, it is illegitimate. Let $\mathcal{D}_{\mathrm{tot}}$ denote the set of all legitimate devices in the network, and let $\mathcal{D}_{\mathrm L} \subset \mathcal{D}_{\mathrm{tot}}$ represent the subset of legitimate devices that are active in a given transmission round. The sizes of these sets are denoted by $|\mathcal{D}_{\mathrm{tot}}| = D_{tot}$ and $|\mathcal{D}_{\mathrm L}| = D_{L}$, respectively. We further define the set of illegitimate devices as $\mathcal{D}_{\mathrm I}$ with cardinality $|\mathcal{D}_{\mathrm I}| = D_{I}$, where $\mathcal{D}_{\mathrm L}\cap \mathcal{D}_{\mathrm I}=\emptyset$. 


RF impairments, arising from unavoidable manufacturing variations, even among devices produced by the same vendor, introduce slight distortions to the transmitted waveforms \cite{11556327,10978824,11369874}. 
Due to device-dependent hardware variations, each device $k$ is associated with a unique set of parameters $\theta_k$ and a corresponding mapping function $\psi_k(\cdot)$ from the ideal baseband signal $s(t)$ to the equivalent baseband signal of the actual transmitted signal under RF impairments, which will be detailed in Section \ref{section Simulation Results}. 
We adopt the OOK modulation scheme in this paper to enable the OHC-based super-sparse spreading. For a binary input symbol $x \in\{0, 1\}$, the ideal OOK baseband signal is given by $s(t) = Axg(t)$, where $A$ denotes the signal amplitude and $g(t)$ is a rectangular pulse with unit amplitude and duration $T_s$. \textcolor{blue}{In narrowband IoT systems, each binary symbol occupies one channel use with a relatively long duration $T_s$, allowing the receiver to collect enough waveform samples within a single active OOK symbol to extract RF fingerprints and support the feasibility of RFFI under single-symbol observation.} 
For analytical tractability, we assume that all devices $k \in \mathcal{D}_{\mathrm{tot}} \cup \mathcal{D}_{\mathrm I}$ follow a common RF impairment model, while their parameters remain device-specific. During the network initialization phase, the BS extracts RF impairment features from all $D_{\mathrm{tot}}$ legitimate devices to form a feature set that is stored in a pre-registered access list. RF features of illegitimate devices are assumed to be unknown to the BS and are not included in the list. The access list is subsequently used for RFFI-enabled message authentication.
\vspace{-1.5em}

\subsection{Transmitters}
\begin{figure}[t!]\centering \vspace{-0em}
	{\epsfig{file=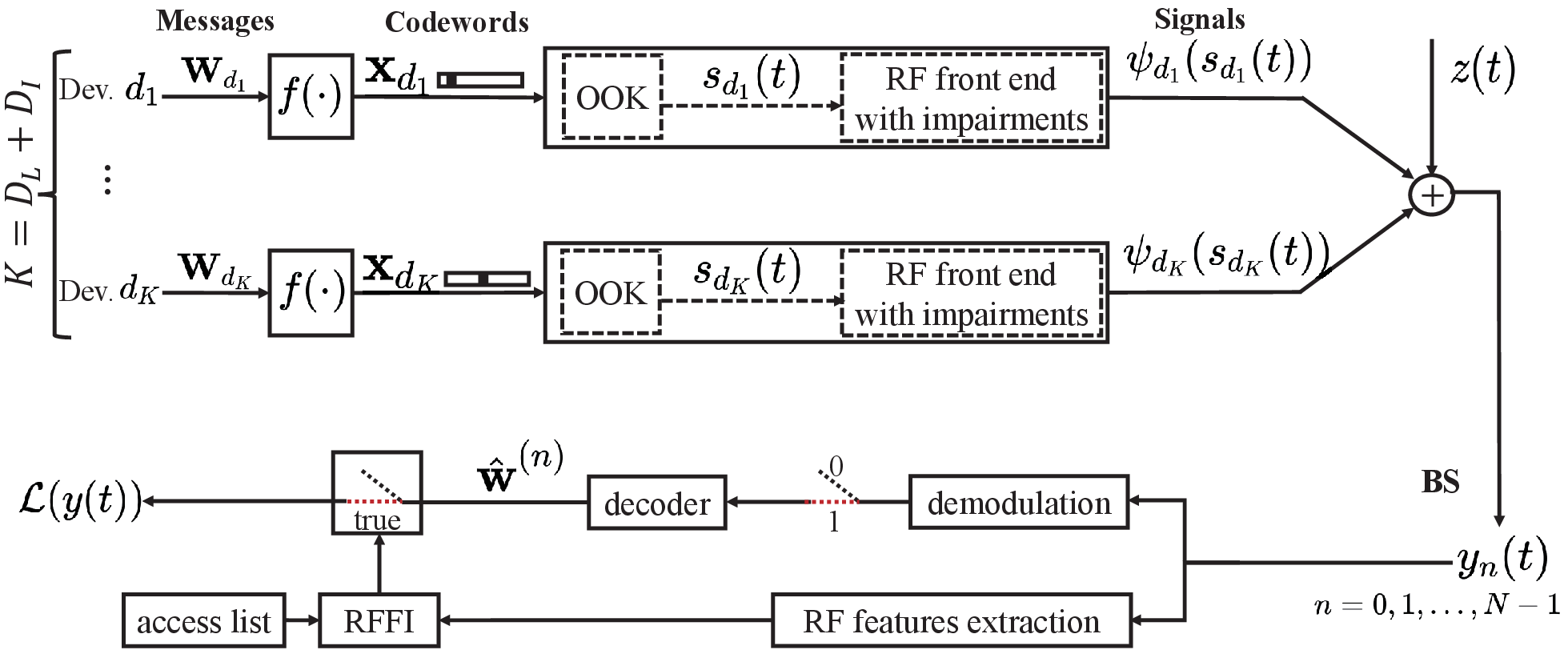, width=0.5\textwidth, clip=}}\vspace{-0.5em}
	\caption{ The overall system block diagram for an uplink transmission round. 
		\vspace{-0em}    }\label{scheme1}   \vspace{-2.5em} 
\end{figure}

As shown in Fig.~\ref{scheme1}, following the URA protocol, all devices share a common codebook, which is equivalently described by the encoding mapping $\mathbf{x} = f(\mathbf{w})$ in the proposed scheme.
Let $\mathbf{w} \in \{0,1\}^B$ denote a $B$-bit message, and let $q \in \{0,\ldots,2^B-1\}$ be its corresponding decimal index. Let $N \triangleq 2^{B}$ denote the number of channel uses. Then $\mathbf{x} \in \{0,1\}^{2^B}$ is a one-hot codeword given by 
\begin{align}  
x_i=
\begin{cases}
1, & \text{if } i=q \text{ the index of } \mathbf{w},\\
0, & \text{otherwise}.
\end{cases}
\label{e1} 
\end{align}
Accordingly, the common codebook matrix can be viewed as an identity matrix. When the codeword $\mathbf{x}$ is applied to the OOK modulator, the resulting ideal baseband signal is denoted by $s(t)$, which represents the concatenation of $N$ OOK symbol waveforms. The transmitted signal is then generated by the RF front end with impairments, which are characterized by the mapping function $\psi(\cdot)$ that maps the ideal baseband signal $s(t)$ to the equivalent baseband signal of the actual transmitted signal. For a given $B$-bit message, only one of the $N$ channel uses, indexed by its decimal representation, is occupied, while the remaining $N-1$ remain idle. In the proposed scheme, we assume that the illegitimate devices are aware of the OHC structure and the OOK modulation scheme.

We now describe an uplink transmission round, shown in Fig.~\ref{scheme1}. In the proposed scheme, each active device, whether a legitimate or an illegitimate device, independently selects its message uniformly at random from the binary sequence space $\{0,1\}^B$, such that each message occurs with equal probability $1/2^B$. Let $ \mathcal{W}_{\mathrm L} = \{ \mathbf{w}_\ell \mid \ell \in \mathcal{D}_{\mathrm L} \} $ denote the set of legitimate messages transmitted by the active legitimate devices, and let $\mathcal{W}_{\mathrm I} = \{ \mathbf{w}_i \mid i \in \mathcal{D}_{\mathrm I} \}$ denote the set of illegitimate messages transmitted by the illegitimate devices. For messages $\mathbf{w}_k, k \in \mathcal{D}_{\mathrm L} \cup \mathcal{D}_{\mathrm I}$, after one-hot encoding $\mathbf{x}_k = f(\mathbf{w}_k)$, the equivalent baseband representation of the actual transmitted signal is given by $\psi_k\left(s_{k}(t)\right)$, where the ideal transmitted signal is
$s_k(t)=A\sum_{n=0}^{N-1}x_{k,n}g(t-nT_s)$.
Here, $x_{k,n}$ is the $n$-th element of the codeword $\mathbf{x}_k$, representing the binary symbol transmitted by device $k$ during the $n$-th channel use. Each transmission block consists of $N$ channel uses. In practice, synchronization errors or timing offsets may introduce inter-channel use interference, which can be mitigated using existing synchronization techniques \cite{8643773}. For analytical tractability, we assume perfect symbol- and block-level synchronization.

\begin{figure}[t!]\centering \vspace{-0.1em}
	 {\epsfig{file=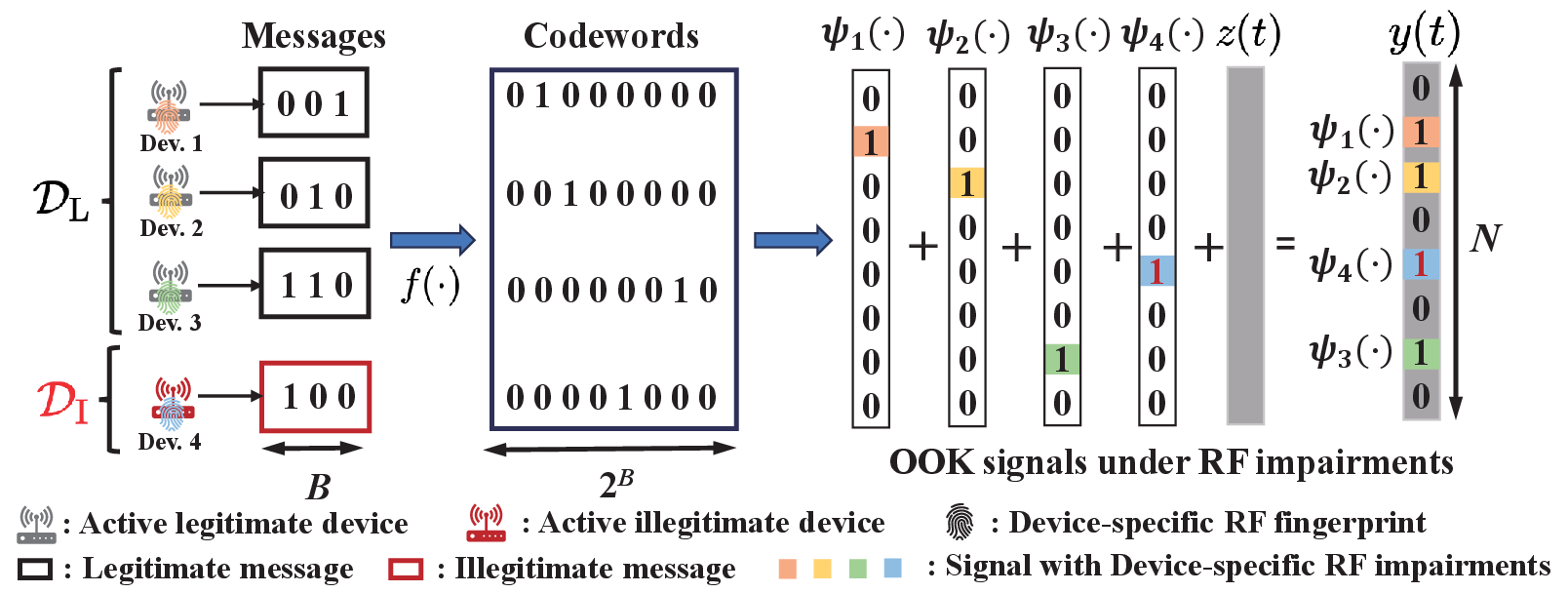, width=0.5\textwidth, clip=}}\vspace{-0.5em}
	\caption{ The proposed OHC-based URA scheme for $B$ = 3. 
		\vspace{-1em}    }\label{scheme2}   \vspace{-1.5em} 
\end{figure}

An example of the transmitted and received signals in the proposed scheme is shown in Fig.~\ref{scheme2} for $B=3$. \textcolor{blue}{Since the codeword length $N=2^B$ grows exponentially with the message length $B$, the code rate $B/2^B$ decreases rapidly as $B$ increases. Therefore, the proposed scheme is suitable for ultra-short-payload scenarios. For example, with $B$ = 12, the corresponding code rate is $\frac{12}{4096}\approx \frac{1}{341}$, which is comparable to those used in existing URA schemes (e.g., $\frac{1}{300}$) \cite{9153051,9797778}.}

\vspace{-1.1em}
\subsection{Receiver}\label{subsec:Receiver} 
The received signal over $N$ channel uses at the BS is the superposition of signals from legitimate and illegitimate devices, yielding
\begin{align}
y(t) = \sum_{k \in (\mathcal{D}_{\mathrm L} \cup\mathcal{D}_{\mathrm I})} \psi_k(s_k(t)) + z(t),  \ t\in[0,NT_s),
\label{e3} 
\end{align}
where $z(t)$ denotes the additive white Gaussian noise with a variance of $\sigma^2$. At the BS, message decoding and authentication are performed on a channel-use basis. Let $y_n(t)=y(t)$ for $ t\in[nT_s,(n+1)T_s)$, $n = 0,1,\dots,N-1$, denote the received signal in the $n$-th channel use. The BS first performs OOK demodulation on each channel use and proceeds with decoding and authentication only when the symbol is detected as active. We consider coherent OOK demodulation with an ideal receiver employing a detection threshold $A/2$ for a transmitted symbol with amplitude $A$, where $y_n(t)$ is demodulated as $x_n = 1$ if the received amplitude exceeds the threshold. For decoding, the estimated message is given by the binary representation of the index $n$, i.e., $\hat{\mathbf{w}}^{(n)} = \mathrm{bin}(n)$.

For authentication, features extracted from $y_n(t)$ are processed via open-set RFFI: a candidate identity is predicted and compared with its registered reference. The message $\hat{\mathbf{w}}^{(n)}$ is accepted if the similarity error is below a threshold, and rejected otherwise \cite{Cai2025OpenSet}. Note that when signal collisions occur during a channel use, the extracted RF impairment features become a mixture of features from multiple devices, in which case we assume authentication fails. Finally, the list of recovered and authenticated messages is given by $\mathcal{L}(y(t)) = \left\{ \hat{\mathbf{w}}^{(n)} \mid n \in \mathcal{S} \right\}$,
where $\mathcal{S} \subseteq \{0,1,\dots,N-1\}$ denotes the set of channel-use indices for which signals are successfully detected and authenticated. \textcolor{blue}{Similar to \cite{9153051,9797778}, we assume that the BS knows the number of active legitimate devices. The receiver sequentially processes the channel uses and retains messages that pass both detection and authentication, terminating once $D_L$ messages have been accepted.}

\vspace{-1em}

\subsection{Performance Metrics} 

In the URA system, the BS aims to recover all the legitimate messages $\mathbf{w}_\ell, \ell \in \mathcal{D}_{\mathrm L}$, based on the received signal $y(t)$,
while preventing messages from illegitimate devices from being accepted. The list of recovered messages is defined as $\mathcal{L}(y(t))$ with $|\mathcal{L}(y(t))| \le D_L$, and the PUPE of active legitimate devices is defined as 
\begin{align}
P_{\mathrm L}=\frac{1}{D_L} \sum_{\ell\in \mathcal{D}_{\mathrm L}}^{}\mathrm{Pr} \left(\mathbf{w}_\ell\notin \mathcal{L}(y(t)) \right). \label{e5}
\end{align}
If a forged message is included in the recovered list, the illegitimate device is considered to have successfully performed spoofing. Defining $P_{\mathrm I}$ as the probability of successful spoofing of a forged message, we have  
\begin{align}P_{\mathrm I}=\frac{1}{D_I} \sum_{i\in \mathcal{D}_{\mathrm I}}^{}\mathrm{Pr} (\mathbf{w}_i\in \mathcal{L}(y(t)) ). \label{e6}\end{align}

\vspace{-1em} 

\section{Performance Analysis }
In this section, we analyze the effects of code rate, $E_b/N_0$, the number of active legitimate devices, the number of illegitimate devices, and the RFFI error probability on the PUPE and successful spoofing probability $P_{\mathrm I}$. 
\vspace{-1em} 
\subsection{Single-Device Collision-Free Channel-Use Analysis }

Recall that the demodulation and authentication are performed on a channel-use basis, as described in Section~\ref{subsec:Receiver}. Accordingly, in the absence of signal collisions, the performance at a single channel use serves as the basis for the system-level analysis.

\subsubsection{OOK Demodulation}
 We assume that the equivalent baseband signal energy of a single OOK symbol is approximately equal to that of the ideal OOK baseband signal, i.e., $E_s=\int_{iT_s}^{(i+1)T_s}\psi(s(t))^2 \mathrm{d}t \approx\int_{iT_s}^{(i+1)T_s}s(t)^2 \mathrm{d}t =A^2T_{\mathrm s}$ for $x_i=1$ according to \eqref{e1}. Nevertheless, the impact of this approximation will be further discussed in Section \ref{section Simulation Results}. 

As described in Section~\ref{subsec:Receiver}, we consider the coherent OOK demodulation. When no signal collision occurs within a single channel use, the miss-detection and false-alarm probabilities are $
P_{1\to0}=P_{0\to1}
=Q\!\left(\sqrt{\frac{E_s}{2N_0}}\right)
=Q\!\left(\sqrt{\frac{B E_b}{2N_0}}\right)
$ \cite{542196}.
The equality on the right-hand side holds due to the OHC structure and OOK modulation, whereby the total energy of the $B$-bit message is concentrated into a single OOK symbol.

\subsubsection{RFFI-Based Authentication}
In a collision-free channel use, a miss detection occurs with probability $P_{\mathrm{md}}$ when a legitimate signal is not authenticated, while a false alarm occurs with probability $P_{\mathrm{fa}}$ when a non-legitimate signal is incorrectly accepted as legitimate. 
In practice, RFFI
is implemented via RF fingerprint (e.g., power amplifier (PA) nonlinearity) feature extraction and machine learning-based
classification techniques \cite{10978824,11369874,105555}. Its performance depends on the extraction
method, classifier, and operating conditions, making closed-form expressions intractable. We therefore model RFFI via $P_{\mathrm{md}}$ and $P_{\mathrm{fa}}$ for theoretical analysis, while adopting the practical parameter ranges reported in \cite{105555} for numerical evaluation.

\vspace{-1em}
\subsection{Multi-Device Analysis over $N$ Channel Uses}
For analytical convenience, we assume that the numbers of active legitimate and illegitimate devices remain constant in each communication round, denoted by $D_L$ and $D_I$, respectively.

\subsubsection{Three Types of Messages}
There are three types of messages in the recovered list $\mathcal{L}(y(t))$. Type-A: legitimate messages transmitted by active legitimate devices. Type-B: illegitimate messages injected by illegitimate devices. Type-C: erroneously decoded messages. We define $P_{A}, P_{B}$, and $P_{C}$ as the probabilities that a given channel use yields a Type-A, Type-B, and Type-C recovered message, respectively.

The probability $P_A$ corresponds to the event that a given channel use yields a Type-A recovered message. This occurs
when exactly one of the $D_L$ active legitimate devices selects that channel use, while the remaining $D_L+D_I-1$ devices do not. Since each device independently selects one of the $N$ channel uses with equal probability, this event occurs with
probability $\binom{D_L }{1} \frac{1}{N}
\left(1-\frac{1}{N}\right)^{D_L+D_I-1}$.
Conditioned on this collision-free event, the transmitted OOK symbol must be correctly demodulated, and the legitimate
signal must be successfully authenticated, which occurs with probabilities $(1-P_{1\to0})$ and $(1-P_{\mathrm{md}})$,
respectively. Therefore, we have
\begin{align}
 P_A= & \frac{D_L}{N}\left (1 - \frac{1}{N} \right ) ^{D_L+D_I-1} \left ( 1-P_{1 \to 0} \right ) \left ( 1-P_{\mathrm{md}} \right ). \label{e9}
\end{align}
A Type-B recovered message occurs when exactly one of the $D_I$ illegitimate devices transmits over a given
channel use, while the remaining $D_L + D_I - 1$ devices do not. Conditioned on this collision-free event, the transmitted
OOK symbol is correctly demodulated, and the signal is erroneously authenticated as legitimate. Therefore, $P_B$ can be written as
\begin{align}
 P_B= & \frac{D_I}{N}\left (1- \frac{1}{N} \right ) ^{D_L+D_I-1} \left ( 1-P_{1 \to 0} \right ) P_{\mathrm{fa}}. \label{e10}
\end{align}
Similarly, a Type-C recovered message occurs when no device transmits over a given channel. Conditioned on this event,
a false alarm occurs in OOK demodulation, and the resulting signal is further falsely authenticated as legitimate by the RFFI module. Therefore, $P_C$ can be written as
\begin{align}
 P_C= & \left (1- \frac{1}{N} \right ) ^{D_L+D_I} P_{0 \to 1}  P_{\mathrm{fa}}. \label{e11}
\end{align}

\subsubsection{$P_{\mathrm L}$ and $P_{\mathrm I}$}
Based on the above analysis, the probability that a given channel use yields a recovered message in $\mathcal{L}(y(t))$ is $P=P_A+P_B+P_C$. Hence, over $N$ channel uses, the expected number of recovered
messages is $PN$. Since the size of the recovered message list $\mathcal{L}(y(t))$ is limited to at most $D_L$, we approximate its expected size as $\min(PN, D_L)$ for analytical tractability. 

Among all recovered messages, the fraction of Type-A messages is $P_A/P$. Therefore, the expected number of correctly
recovered legitimate messages is approximated by $\frac{P_A}{P}\min(PN,D_L)$. Accordingly, the PUPE of active legitimate devices can be rewritten as
\begin{align}
P_{\mathrm L}=1-\frac{1}{D_L}\frac{P_A}{P}\min(PN,D_L).  \label{e12}
\end{align}
Similarly, the fraction of Type-B messages among all recovered messages is $P_B/P$. Therefore, the expected number of
illegitimate messages in the recovered list $\mathcal{L}(y(t))$ is approximated by $\frac{P_B}{P}\min(PN,D_L)$.
Accordingly, the average successful spoofing probability per illegitimate device can be
written as
\begin{align}
P_{\mathrm I}=\frac{1}{D_I}\frac{P_B}{P}\min(PN,D_L). \label{e13}
\end{align}

\textcolor{blue}{We also analyze the asymptotic performance of $P_{\mathrm L}$ by considering the regime where $E_b/N_0$ tends to infinity. Both $
P_{1\to0}$ and $P_{0\to1}$ tend to zero in this regime. By substituting these into \eqref{e9}, \eqref{e10}, and \eqref{e11}, the asymptotic expressions for $P_{\mathrm L}$ can be derived as
\begin{align}P_{\mathrm L}^{\mathrm a}\!\!=\!\!
\begin{cases}
1\!-\!(1\!-\!P_{\mathrm{md}})(1-\frac{1}{N})^{D_L+D_I-1}, & \text{if } PN < D_L,\\
\frac{D_I P_{\mathrm{fa}}}{D_L(1-P_{\mathrm{md}})+D_I P_{\mathrm{fa}}} , & \text{otherwise}.
\end{cases} \label{e14}
\end{align}}

\vspace{-1em}
\section{Numerical Results} \label{section Simulation Results}

As reported in \cite{105555}, $P_{\mathrm{md}}$ and $P_{\mathrm{fa}}$ for RFFI with OOK signals lie in $[0, 0.02]$, which we use below to evaluate the minimum required $E_b/N_0$, $P_{\mathrm{L}}$, and $P_{\mathrm{I}}$. We note that the resulting system-level authentication performance is conditioned on this assumed RFFI operating point, and would vary with the specific RFFI implementation adopted.

\begin{figure}[b!]\vspace{-2.2em} \centering 
	 {\epsfig{file=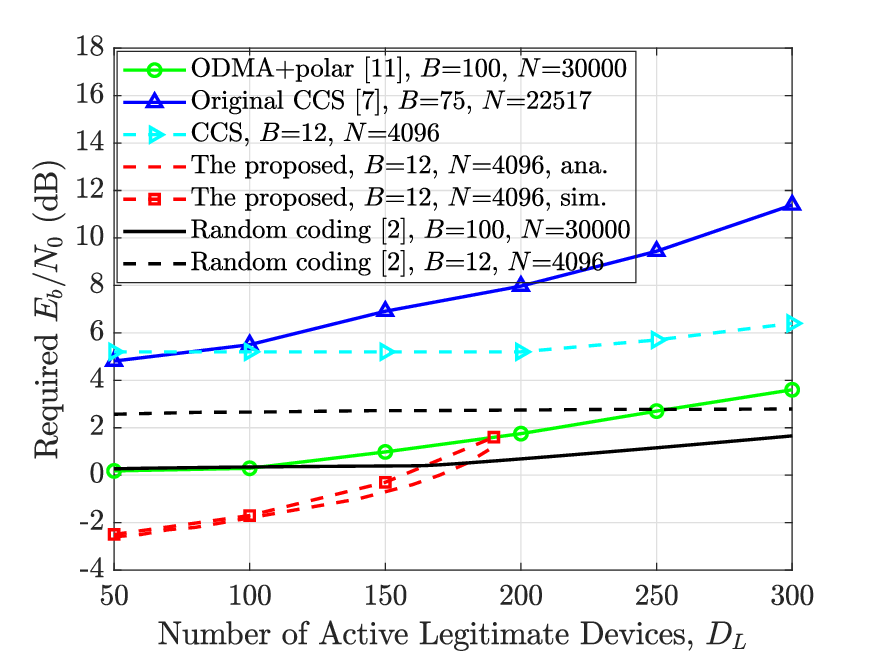, width=0.38\textwidth, clip=}}\vspace{-0.5em}
	\caption{Required $E_b/N_0$ for PUPE = 0.05 vs. the number of active legitimate devices $D_L$ for $P_{\mathrm{md}}=P_{\mathrm{fa}}=0$ \textcolor{blue}{and $D_I=10$}. 
	   }\label{fig1} 
       \vspace{-0em}
\end{figure}

In Fig.~\ref{fig1}, we compare the required $E_b/N_0$ of the proposed scheme with those of existing URA schemes in \cite{9153051,10415634} and the achievable bound in \cite{8006984}. Although the proposed scheme supports a limited number of active users, it can effectively mitigate spoofing attacks under ideal RFFI conditions, thereby enabling secure transmission in URA scenarios. In addition, since the proposed scheme adopts the OHC-based extremely sparse-spreading strategy with ${E_s}/{N_0}={BE_b}/{N_0}$, it requires a lower $E_b/N_0$ than the matched CCS baseline with $B=12$ and $N=4096$ at the same ${E_s}/{N_0}$. This is because the CCS baseline adopts the dense spreading structure with ${NE_s}/{N_0}={BE_b}/{N_0}$.

\textcolor{blue}{To characterize the impact of waveform distortion caused by RF impairments, we also
conduct $5000$ Monte Carlo trials for each simulation point}, adopting PA nonlinearity \cite{10930706} as the RF impairment model without RF impairment compensation, i.e., $\psi(s(t))=\frac{\alpha s(t)}{1+\beta s(t)^2}$, with device parameters uniformly
distributed within $\pm 5\%$ of the default values $[\alpha=2.1587,\beta=1.1417]$. Since compensation is not considered, RF impairments slightly degrade OOK demodulation, causing a minor gap between analytical and simulation results.

\begin{figure}[t!]\centering \vspace{-0.6em}
	 {\epsfig{file=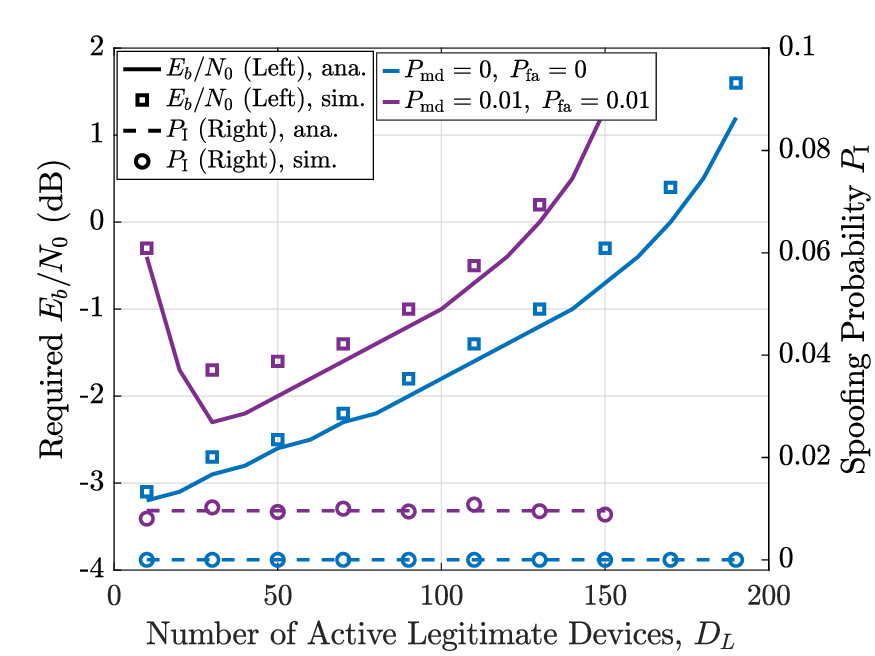, width=0.38\textwidth, clip=}}
	\caption{ Required $E_b/N_0$ and $P_{\mathrm I}$ for PUPE = 0.05 vs. the number of active legitimate devices $D_L$ for $B=12$, $D_I=10$, and $N=4096$.
		    }\label{fig2}
            \vspace{-1em}
\end{figure}

In Fig. \ref{fig2}, the left y-axis illustrates the minimum required $E_b/N_0$ of the proposed scheme under both imperfect and perfect RFFI modules. It can be observed that the imperfect RFFI module reduces the number of supportable active legitimate devices and increases the minimum required $E_b/N_0$. In addition, according to \eqref{e12}, we obtain $PN=D_L$ when $D_L=28$ through numerical calculations for the imperfect RFFI module case (i.e., $P_{\mathrm {md}}=P_{\mathrm {fa}}=0.01$). \textcolor{blue}{When $PN<D_L$, the number of recovered messages is smaller than the number of active legitimate devices, and the PUPE is primarily determined by $P_{A}$. When $PN\geq D_L$, the PUPE is mainly affected by $P_{B}$ and $P_{C}$.}  Therefore, when $D_L$ is greater than or smaller than $28$, $P_{\mathrm L}$ in \eqref{e12} operates in different regimes, resulting in different trends in the minimum required $E_b/N_0$. The right y-axis represents the spoofing probability $P_{\mathrm I}$ under the constraint $P_{\mathrm L} = 0.05$. It can be observed that the imperfect RFFI module also increases $P_{\mathrm I}$, since $P_{\mathrm I}$ is affected by $P_{\mathrm{md}}$ and $P_{\mathrm{fa}}$ according to \eqref{e13}.

\begin{figure}[t!]\centering \vspace{-0.3em}
 	 {\epsfig{file=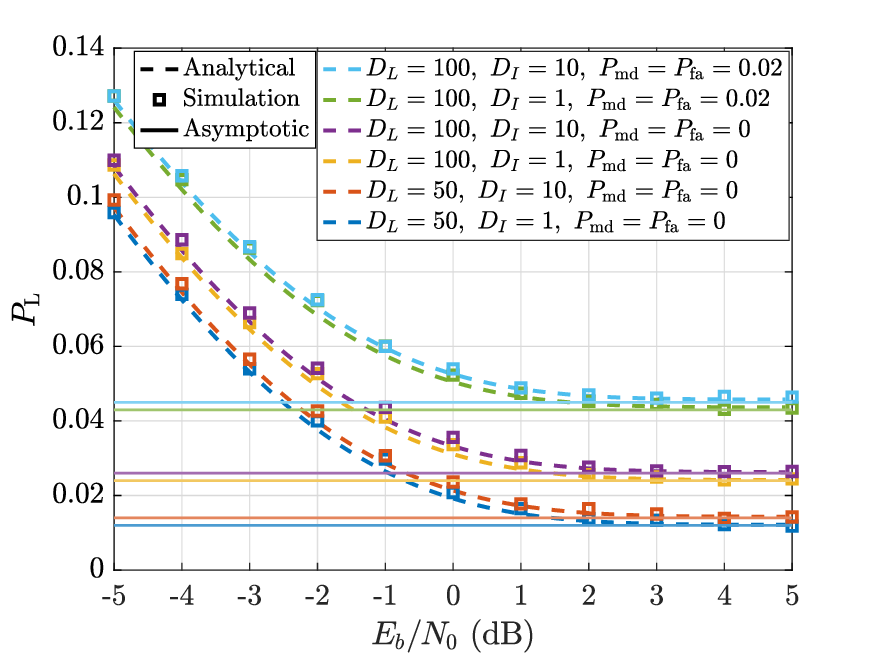, width=0.38\textwidth, clip=}}\vspace{-0.5em}
 	\caption{ PUPE vs. $E_b/N_0$ for $B=12$, $N=4096$.
 		  }\label{fig3}
           \vspace{-1.5em}
 \end{figure}
\textcolor{blue}{Fig.~\ref{fig3} plots the impact of $D_L$, $D_I$, $E_b/N_0$, and imperfect RFFI on the PUPE performance. As $E_b/N_0$ increases, we observe that the PUPE curves converge to an error floor. Specifically, $D_L$ and RFFI imperfections dominate the error floor $P_{\mathrm L}^{\mathrm a}$ \eqref{e14}, whereas the impact of a small $D_{I}$ is relatively marginal. These observations indicate that, in the high $E_b/N_0$ regimes, the PUPE performance is mainly limited by message collisions and RFFI imperfections rather than additive noise as analyzed in \eqref{e14}.}

\vspace{-1.5em}
\section{Conclusions}
This paper has proposed an OHC-based URA scheme that integrates RFFI for message authentication. 
While the current design has been tailored to ultra-short payload scenarios, it introduces an authentication stage that can be combined with existing URA schemes to accommodate higher payloads. \textcolor{blue}{For example, in two-phase URA systems, the proposed scheme can be used for the first-phase message. In segmentation-based URA systems, it can be used for one selected block, while the remaining blocks follow conventional CS-based transmission.}


\linespread{1}
\vspace{-1em}
\bibliographystyle{IEEEtran}
\bibliography{WCL}

\end{document}